\documentclass[final,5p,times,twocolumn]{elsarticle}

\usepackage{amssymb}
\usepackage{color}

\journal{Physics Letters B}

\begin{document}




\title{Rotational excitations in near neutron-drip line nuclei:
       the birth and death of particle-bound rotational bands and the extension
       of nuclear landscape beyond spin zero neutron drip line.}


\author[msu,ku]{A.\ V.\ Afanasjev}

\corref{cor1}
\ead{Anatoli.Afanasjev@gmail.com} 
\cortext[cor1]{Corresponding author}

\address[msu]{Department of Physics and Astronomy, Mississippi State
University, MS 39762, USA}

\address[ku]{Yukawa Institute for Theoretical Physics, Kyoto University,
             Kyoto 606-8502, Japan}

\author[ku]{N.\ Itagaki}

\author[msu,msu1]{D.\ Ray}
\address[msu1]{Institute for Systems Engineering Research (ISER), Mississippi State University, MS 39762, USA}

\begin{abstract}
    Two new mechanisms active in rotating nuclei located in the vicinity of 
neutron drip line have been discovered. Strong Coriolis interaction 
acting on high-$j$ orbitals transforms particle-unbound (resonance) 
nucleonic configurations into particle-bound ones with increasing angular 
momentum. The point of the transition manifests the birth of particle-bound
rotational bands. Alternative possibility of the transition from particle-bound 
to resonance rotational band (the death of particle-bound rotational 
bands) with increasing spin also exists but it is less frequent in the calculations.
The birth of particle-bound rotational bands provides a  mechanism for 
the  extension of nuclear landscape  to neutron numbers which are larger
than those of the neutron drip line in non-rotating nuclei.

\end{abstract}

\maketitle

 One of most fundamental questions in nuclear physics is what 
combinations of protons and neutrons form a nucleus as a bound 
system.  For a given proton number, there is a maximum number 
of neutrons beyond which the formation of bound nuclear systems
is impossible. This limit is known as a neutron drip line.
Although significant experimental efforts have been dedicated to 
the investigation of very neutron rich nuclei \cite{40Mg-42Al,60Ca-PRL.18}, 
the neutron drip line is definitely deliniated only up to $Z=10$. Future 
experimental facilities such as FRIB and FAIR will hopefully extend 
experimental neutron drip line to higher $Z$ up to mass number $A\sim 70$ (see Ref.\ 
\cite{FRIB.2009} and Fig.\ 1 in Ref.\ \cite{AARR.15}). 
These experimental efforts are accompanied by a significant amount of theoretical 
investigations of  neutron-rich limit of the nuclear landscape (see Refs.\ 
\cite{Eet.12,AARR.13,AARR.14} and references quoted therein). They provide 
the predictions for the position of neutron-drip line extracted from binding energies 
of non-rotating ground states. Such line is denoted here as the $I=0$ neutron drip 
line.

   On the other hand, the questions of possible existence of 
particle-bound nuclear states in the nuclei located beyond the $I=0$ neutron 
drip line and physical mechanism leading to such states have not even 
been raised in the literature.  For the first time we show that nucleonic
configurations which are particle-unbound at zero or low spins can 
be transformed into particle-bound ones by collective rotation of
nuclear systems. This leads to new features of rotational bands 
which are distinctly different from those seen before in known nuclei 
and which are discussed in the present manuscript for the first time. 
These features  facilitate the extension of nuclear landscape to higher 
neutron numbers beyond those seen in the $I=0$ neutron drip line.
  
    Static pairing correlations are present in atomic nuclei at spin zero. However, 
a reliable extrapolation of their properties towards neutron drip line still remains a 
challenge. As discussed in Ref.\ \cite{AARR.15}, even a careful fitting of the 
pairing force in known nuclei to experimental odd-even mass staggerings will 
not necessarily lead to a pairing force with a reliable predictive power towards 
the two-neutron drip line. Indeed, there is a substantial difference in the model 
predictions for pairing properties near two-neutron drip line which depends both 
on underlying functional and on the type of employed pairing interaction  (see 
Ref.\ \cite{AARR.15,PMSV.13,YS.08}). For example,  relativistic Hartree-Fock calculations 
with separable pairing predict either similar or somewhat larger  pairing in the 
nuclei near two-neutron drip line as compared with known nuclei (see Fig.\ 2 in 
Ref.\ \cite{AARR.15}). On the contrary, the Skyrme DFT calculations for 
spherical nuclei show the reduction of neutron pairing towards the neutron drip line 
\cite{PMSV.13}.

\begin{figure}[ht]
\centering
\includegraphics[width=8.5cm]{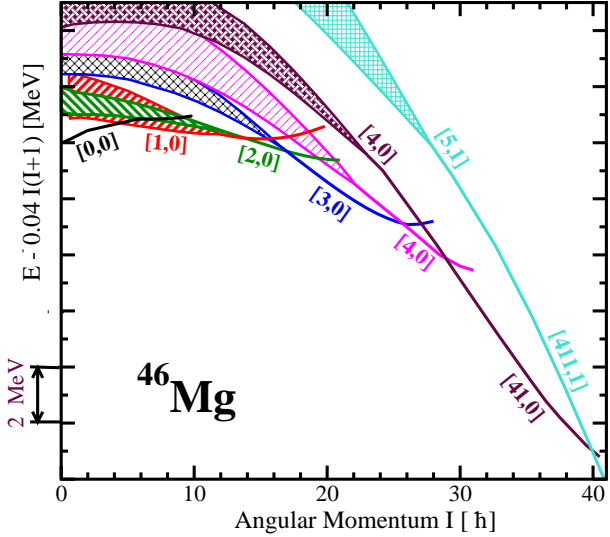}
\caption{Excitation energies of calculated configurations 
in $^{46}$Mg relative to a rotating liquid drop reference $AI(I + 1)$, 
with the inertia parameter $A = 0.04$.  Shaded regions display the
resonance parts of the bands; their height in vertical direction 
schematically illustrates  the width of resonance states and their 
evolution with spin.
}
\label{E_RLD}
\end{figure}

However, the role of pairing is expected to be substantially reduced with 
increasing spin because of Coriolis anti-pairing \cite{RS.80,VDS.83} 
and blocking \cite{RS.80} effects. The angular momentum in the ground state 
bands of the light nuclei is very limited (see discussion in Ref.\ \cite{RA.16}); 
higher spin configurations are built by particle-hole excitations with many of 
them triggering the reduction of pairing by blocking effect. Moreover, the pairing 
drastically decreases after paired band crossings in the proton and neutron 
subsystems\footnote{The treatment of pairing correlations in rotating nuclei requires the methods of 
approximate (such as Lipkin-Nogami method \cite{L.60,N.64}) particle number projection 
\cite{GBDFH.94,SWM.94,CRHB,AER.00}. While their contribution is very important for the 
description of rotational properties at low and moderate spins (see Refs.\ \cite{GBDFH.94,SWM.94,CRHB,AER.00}), 
the role of pairing correlations is  substantially reduced at high spin after paired band crossings in 
proton and neutron subsystems or for the configurations with blocked particles
so that the results obtained in the calculations without pairing become a very good approximation to 
the ones calculated with the LN method \cite{AF.05,RA.16,72Kr-exp}. Note that the configurations 
considered in the present manuscript belong to a such high spin range.
}  \cite{AF.05}. As a consequence, it is expected that pairing 
correlations will have only a minor impact on rotational and deformation 
properties of the nuclei of interest  at spins above $I\sim 10\hbar$. 
Weakening of pairing with angular momentum also leads to a significant
reduction of the coupling with continuum. Note that 
neutron subsystem of Mg nuclei under study is very similar to the one of 
the $N\sim Z$  $A=40-50$ nuclei and in these nuclei pairing plays a negligible 
role above $I\sim 10\hbar$ \cite{RA.16}.  

   Because of these reasons the pairing correlations are neglected in the present 
exploratory study which is focused on medium and high spin states. The calculations 
are performed in the framework of cranked relativistic mean field (CRMF) theory 
\cite{KR.89,KR.93,AKR.96}\footnote{So far only few manuscripts have been 
dedicated to the study of rotating near neutron-drip line 
nuclei. Few ground state rotational bands of light nuclei have been studied in 
Skyrme DFT  framework \cite{NDMMS.01} and in qualitative model of Ref.\ \cite{YS.08}. 
Rotating  cluster configurations of the C neutron-rich isotopes were investigated in the
CRMF framework (Ref.\ \cite{ZIM.15}). Note that pairing correlations have been neglected 
in the studies of Refs.\ \cite{NDMMS.01,ZIM.15}.}. It represents the 
realization of  covariant density functional theory (CDFT)  for rotating nuclei with 
no pairing correlations in one-dimensional cranking approximation \cite{VALR.05}. 
It has been successfully tested in a systematic way on the properties of different 
types of rotational bands in the regime of weak pairing such as normal-deformed, 
superdeformed and smooth terminating bands as well as the bands at the 
extremes of angular momentum (see Refs.\  \cite{AF.05,VALR.05} and references therein).  
In the current study, we restrict ourselves to reflection symmetric shapes, 
which are  dominant deformed shapes  in nuclear chart.

  The properties of neutron-rich nuclei are frequently studied in spherical shell
motivated models (see, for example, Refs.\ \cite{CNPR.98,OSHJSY.10}). However, 
because of numerical reasons such models cannot handle the configurations built 
on intruder orbitals which are involved in the structure of all high spin states.
This is seen for example in the $A\sim 60$ $N\sim Z$ region
in which  highly- and superdeformed configurations are based on intruder orbitals.
Such configurations are successfully described in the cranked versions of relativistic 
and  non-relativistic density functional theories \cite{A60,56Ni.exp-HighD} but they are 
outside the applicability range of spherical shell motivated models (see, for example,
Ref.\ \cite{TOSHU.14}).

    As discussed in detail below, many rotational bands  in near neutron-drip
line nuclei are built from resonance and particle-bound parts. In resonance part of the 
band, at least one single-particle state forming nucleonic configuration is embedded 
into continuum. On the contrary, all single-particle states have negative energy 
(are particle-bound) in particle-bound part of the rotational band. The CRMF theory 
uses basis set expansion for the solution of  the underlying equations. Note that in 
the case of no pairing this is very good  approximation for particle-bound parts of 
rotational bands. This is because  all single-particle 
orbitals have negative energy and there is no coupling with continuum in the case 
of no pairing. Thus, particle-bound parts of rotational  bands are properly described 
using basis set expansion method.  This method would also be appropriate for the 
case when pairing is present but the Fermi level is located reasonably far away from 
continuum threshold\footnote{Note that all systematic state-of-the-art calculations 
of the position of the two neutron-drip line in the density functional and mean field theories 
have been performed in  the computer codes employing basis set expansion 
in large basises (see Refs.\  \cite{Eet.12,AARR.14,RAP.15} and references therein).}. This is 
indeed a case for a number of bands, the Fermi level of which is located at the energy 
$e_F \sim -2$ MeV for top parts of these bands (see discussion below).

    The use of the basis set expansion is more questionable for the nucleonic
configurations in which at least one neutron single-particle orbital is embedded 
in the continuum. This is because of possible couplings with continuum.  As a result, 
the detailed properties of resonance parts of rotational bands are not subject of this
study. Their investigation requires cranking codes formulated in coordinate 
space which to our knowledge  are not available nowadays for density functional
theories.  It is however interesting to mention that in the present CRMF calculations the
resonance and particle-bound parts of rotational band do form smooth rotational
band without sharp changes on the transition from one part to another.

\begin{figure}[ht]
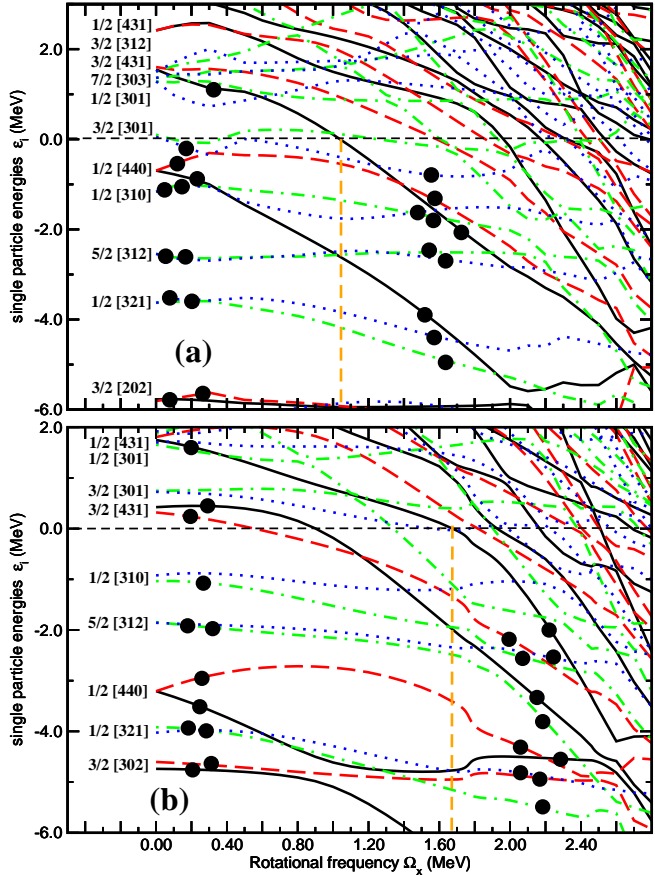

\centering
\includegraphics[width=8.5cm]{fig-2-a.eps}
\includegraphics[width=8.5cm]{fig-2-b.eps}
\caption{Neutron single-particle energies (routhians) 
in the self-consistent rotating potential of $^{46}$Mg as a function of 
rotational frequency $\Omega_x$. They are given along the deformation 
paths of the [3,0] (panel (a)) and combined [5,1]/[411,1] (panel (b))
configurations. Long-dashed  red, solid black, dot-dashed green, and 
dotted blue lines indicate $(\pi = +, r = +i)$, $(\pi = +, r = -i)$, 
$(\pi = -, r = +i)$, and $(\pi = -, r = -i)$ orbitals, respectively. At $\Omega_x = 0.0$ MeV, 
the single-particle orbitals are labeled by the asymptotic quantum 
numbers $\Omega[Nn_z\Lambda]$ (Nilsson quantum numbers) of the dominant 
component of the wavefunction. Solid circles indicate occupied orbitals in
resonance and particle-bound parts of  respective configurations.  Vertical orange 
dashed line indicates the frequency at which the configuration becomes particle bound.}
\label{routh}
\end{figure}

 The CRMF calculations have been performed with the NL3* functional \cite{NL3*}
which is state-of-the-art functional for nonlinear meson-nucleon coupling model
\cite{AARR.14}. It is globally tested for ground state observables in even-even 
nuclei \cite{AARR.14,AA.16}. The CRMF and cranked relativistic Hartree-Bogoliubov 
calculations with this functional provide a very successful description of  different types of rotational bands 
\cite{NL3*,ASN.12,AO.13} both at low and high spins.
    
    The CRMF equations are solved in the basis of an anisotropic three-dimensional 
harmonic oscillator in Cartesian coordinates characterized by the deformation 
parameters $\beta_0$ and $\gamma$ and oscillator frequency 
$\hbar \omega_0 = 41 A^{-1/3}$ MeV
for details see Refs.\ \cite{AKR.96,KR.89}. 
The truncation of basis is performed in such a way that all states belonging 
to the major shells up to $N_F=18$ fermionic shells for the Dirac spinors and 
up to $N_B=20$ bosonic shells for the meson fields are taken into account. 
This truncation scheme provides very good numerical accuracy for the 
nuclei under study (see discussion of Fig.\ \ref{convergence} below).
    
\begin{figure*}[ht]
\centering
\includegraphics[width=12.0cm]{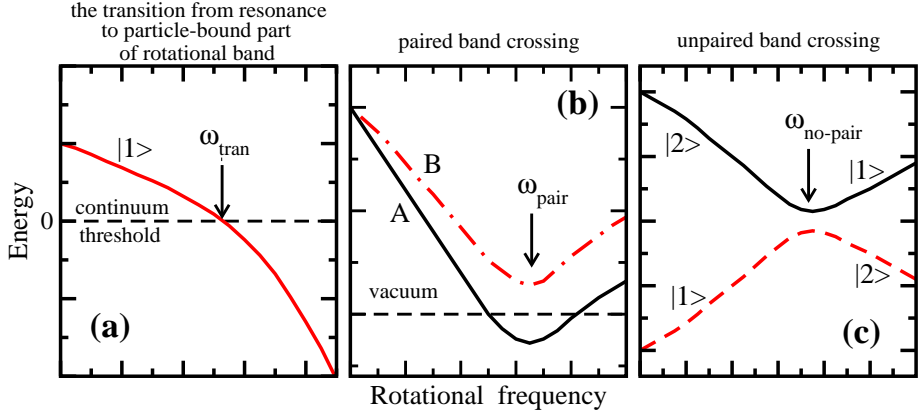}
\caption{Schematic comparison of the transition from resonance to particle-bound part
of rotational band in near neutron-drip line nuclei [panel (a)] with paired [panel (b)]
and unpaired [panel (c)] band crossings in rotating nuclei.  Panel (a) shows the evolution 
of the highest single-particle level occupied in nucleonic configuration with respect 
of continuum threshold leading to the transition from resonance to particle-bound part
of rotational band. The energy evolution of the lowest two quasiparticle orbitals of opposite 
signature with respect of quasiparticle vacuum leading to paired band crossing at $\omega_{pair}$ 
is illustrated in panel (b) (see Sect. 2.5 of Ref.\ \cite{Szy-book} and Sect.\ 5 of Ref.\ \cite{VDS.83} for more 
detailed discussion of paired band crossings).  In panel (c) illustrating unpaired band crossing, the lowest 
(highest) in energy single-particle orbital is occupied (empty). These orbitals exchange their character 
at the crossing frequency $\omega_{no-pair}$ (see Sect. 2.3.5 in Ref.\ \cite{Szy-book}  and
Sect. 12.4 in Ref.\ \cite{NilRag-book}  for more detailed discussion of unpaired band 
crossings). Single-particle orbitals are labelled by $|1>$ and $|2>$ for the case of no pairing.
The conventional labelling of quasiparticle orbitals by letters A, B, ... (see Ref.\ \cite{BF.79-2}) is used 
here.
}
\label{compar}
\end{figure*}

\begin{figure*}[ht]
\centering
\includegraphics[width=17.8cm]{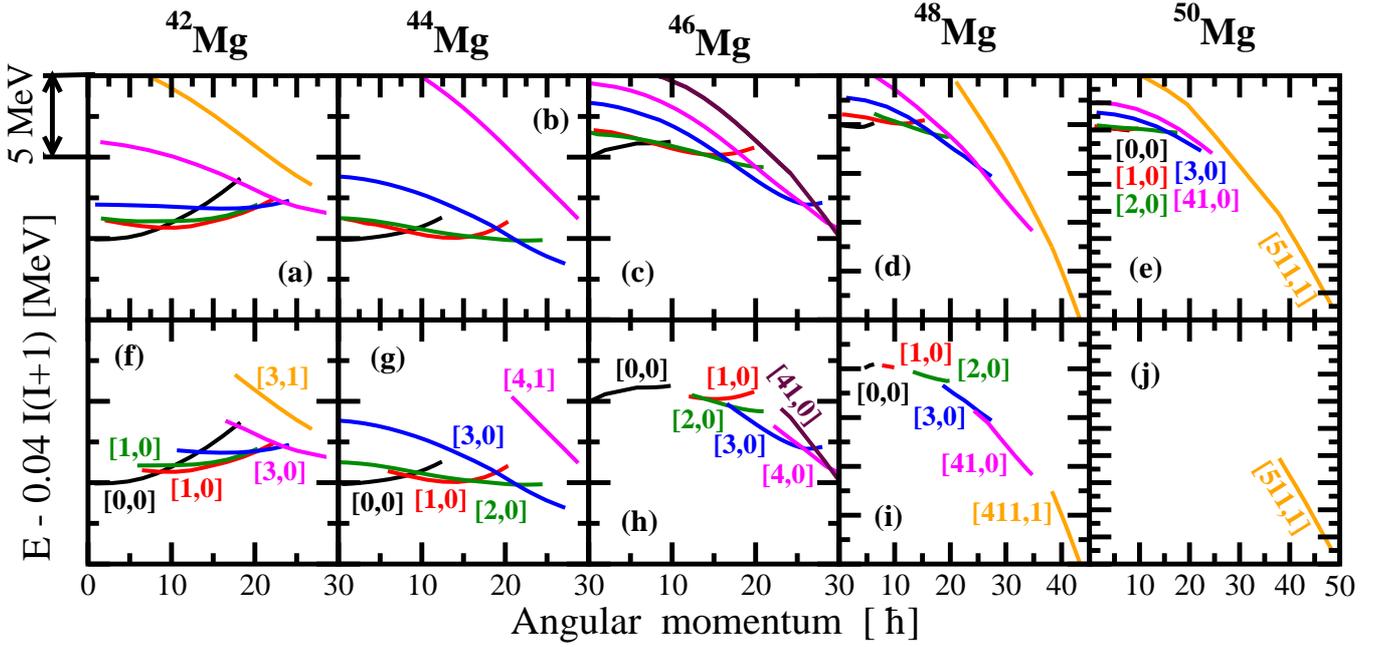}
\caption{(Color online) Yrast and near-yrast rotational bands in even-even Mg nuclei
         close to two-neutron drip line. The $^{46}$Mg nucleus in 
         the middle of figure corresponds to last bound nucleus 
         at spin zero. The $^{48,50}$Mg nuclei are unbound in the
         RHB calculations of Ref.\ \cite{AARR.13,AARR.14} at spin 
         zero. Top panels show both the resonance and particle-bound
         parts of rotational bands. Only particle-bound parts of rotational 
         bands are shown in bottom panels. Note that the energy and spin
         ranges are enlarged for the $^{48,50}$Mg nuclei as compared 
         with lighter nuclei. In all panels, the distance between large ticks
         on vertical axis is equal to 5 MeV.     
         }
\label{systematics}
\end{figure*}

  The rotational structures are studied in even-even $^{42,44,46,48,50}$Mg
nuclei. Their  nucleonic configurations are specified in the calculations by the number of 
proton/neutron orbitals of four combinations of parity $\pi$ and signature 
$r$ occupied from the bottom of the potential. Because the pairing correlations 
are  neglected, the intrinsic structure of the configurations of interest can be 
described by means of dominant single-particle components of  
intruder/hyperintruder/megaintruder orbitals occupied. Thus, the 
calculated configurations are labeled by shorthand 
[$n_{1}$($n_{2}$)($n_{3}$),$p_{1}$] labels, where $n_{1}$, $n_{2}$ 
and $n_{3}$ are the number of neutrons in the $N=4$ intruder, $N=5$ 
hyperintruder and $N=6$ megaintruder orbitals and $p_{1}$ is the 
number of protons in the $N=3$ intruder orbitals. The $n_{3}$ and $n_{2}$ 
numbers are omitted from shorthand labels when respective orbitals are not 
occupied.  Note that the configurations which are yrast or near-yrast in the 
$I\leq 30\hbar$ range are shown in Figs.\ \ref{E_RLD} and \ref{systematics} only 
up to the spin of first unpaired band crossing with the change of at least one label 
(either $n_{1}$ or $n_{2}$ or $n_{3}$). On the contrary, higher spin configurations 
are shown with such band crossings included with respective labelling of the before/after 
band crossing  branches of the configuration.

  According to relativistic Hartree-Bogoliubov calculations of
Ref.\ \cite{AARR.14}, the $^{46}$Mg nucleus with 12 protons and 
34 neutrons is last neutron bound nucleus in the NL3* 
functional\footnote{There are substantial uncertainties in the
predictions of the position of two-neutron drip line in the Mg 
isotopes (see Refs.\ \cite{Eet.12,AARR.13,AARR.14,60Ca-PRL.18}) 
which is typically located from $^{40}$Mg up to $^{46}$Mg. However, 
available experimental data does not allow to discriminate
between these predictions (see Refs.\ \cite{40Mg-42Al,60Ca-PRL.18}).}.
The configurations of this nucleus, which appear as yrast or 
near-yrast at least in some spin range, are shown in Fig.\ 
\ref{E_RLD}. Its ground state band has the [0,0] configuration 
in the CRMF calculations. The 
rotational states of this band are neutron bound since all 
occupied neutron single-particle routhians have negative 
energy.

\begin{figure}[ht]
\centering
\includegraphics[width=8.9cm]{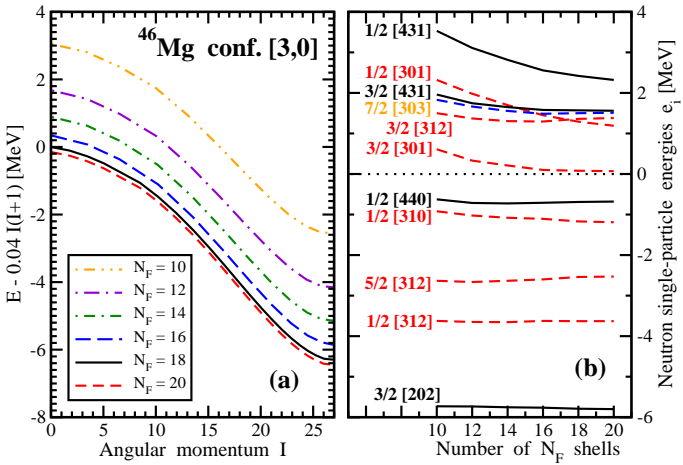}
\caption{The dependence of calculated binding energies $E$ (panel (a)) and 
neutron single-particle energies $e_i$ (panel (b)) on the number of fermionic
shells $N_F$ included in the calculations. The results of the calculations 
for the [3,0] configuration in $^{46}$Mg are presented here. Panel (a) shows 
binding energies with respect of a rotating liquid drop (RLD) reference 
[$E_{RLD}=A(I(I+1)$] with the same  inertia parameter $A$ as that used in Fig.\  
\ref{E_RLD}.  Note that the energy of the $E - E_{RLD}$ curve calculated with 
$N_F=18$ is normalized to zero at $I=0$. Panel (b) shows single-particle energies 
at no rotation; their evolution as a function of  rotational frequency is displayed 
in Fig.\ \ref{routh}a for the $N_F=18$ case. Dotted line on panel (b) shows continuum 
threshold. }
\label{convergence}
\end{figure}

 The angular momentum in the valence space configurations is limited  
 (see Ref.\ \cite{RA.16}). Thus, higher spin states are build by means of particle-hole 
excitations to intruder and hyperintruder orbitals which increase both the 
maximum spin within the configurations and quadrupole deformation of the 
configurations (see discussion in Ref.\  \cite{RA.16} and Fig.\ \ref{E_RLD}).
Fig.\ \ref{E_RLD} clearly shows that with increasing spin the 
yrast line of the $^{46}$Mg nucleus is gradually built by the 
configurations which have larger and larger intruder/hyperintruder 
content  reaching [411,1] at the highest calculated spins.

This process leads to new physical mechanism which is not active in rotational 
bands of the nuclei in the vicinity of the $\beta$-stability line. This mechanism is 
best illustrated on the example of the [3,0]  configuration the neutron routhian 
diagram of which is shown in Fig.\  \ref{routh}(a). In this configuration, the 
$3/2[431](r=-i)$ orbital is the highest in energy occupied positive parity intruder 
orbital.  Note that in the calculations without pairing, the energy of the highest 
occupied neutron orbital corresponds  to the energy of neutron Fermi level
\footnote{The stability of the element with respect of one-neutron decay is defined
by one-neutron separation energy $S_n$. Note that in the calculations 
with pairing,  $S_n \approx -\lambda_n + \Delta_n$ 
in even $N$ nuclei with $\lambda_n$ and  $\Delta_n$ being the neutron Fermi 
energy and pairing gap, respectively. In the case of no pairing, 
$S_n \approx -\lambda_n$ with the energy of last  occupied neutron 
orbital equal to $\lambda_n$. Thus, the nucleonic configuration the energy
of the last occupied neutron orbital of which is negative is expected to
be particle bound.}  \cite{RS.80, NilRag-book,PhysRep-SBT}. At rotational frequency $\Omega_x <1.03$ MeV, 
the $3/2[431](r=-i)$ orbital is particle unbound since its single-particle energy is 
positive. Above this frequency, its energy becomes negative and this orbital dives 
deeper into nucleonic potential with increasing rotational frequency. Above 
$\Omega_x\sim 2.0$ MeV, the  structure of the [3,0] configuration changes 
because of the occupation of the lowest $N=5(r=+i)$ orbital (see Fig.\ \ref{routh}(a)).

   Fig.\  \ref{E_RLD} shows the [3,0] configuration from low spin up to the 
spin at which the configuration change takes place.  At low spin (up to 
$I\sim 16\hbar$), this rotational band can exist only as a band embedded 
in particle continuum (see Refs.\ \cite{GEF.13,FNJMP.16}) (further 
'resonance band') since last occupied orbital, namely, the $3/2[431](r=-i)$ 
orbital is particle unbound. The resonance width  of the states in the  resonance 
band is energy dependent  and its existence depends on whether respective 
neutron  or other decay channels are open or closed (see Refs.\  \cite{GEF.13,FNJMP.16}). 
Because of the location in continuum, the detailed properties of resonance bands or the 
question of their existence are not the subjects of the present study.  However, at spin 
$I > 16 \hbar$  (at $\Omega_x \geq 1.03$ MeV), the configuration [3,0]  becomes particle 
bound. Thus, respective rotational band  changes its character from 
particle-unbound resonance band (at $I<16 \hbar$)  to particle-bound 
rotational band  (at $I>16\hbar$) with discrete rotational states  of extremely 
narrow width.
        
   So the transition from resonance part to particle-bound part of rotational band
in the [3,0] configuration is defined solely by the evolution in energy of the $3/2[431](r=-i)$ 
orbital with respect of the continuum threshold as a function of rotational
frequency (see Fig.\ \ref{compar}a). This transition takes place when the single-particle 
state crosses the continuum threshold at frequency $\omega_{tran}$.   The question 
is whether cranking description in rotating frame provides a proper reproduction 
of energy evolution of the single-particle state and its crossing of continuum threshold. 
It turns out that the situation discussed above and shown in Fig.\ \ref{compar}a
is extremely similar to the description of paired and unpaired band crossings in rotating 
nuclei.   In cranking model description of rotating nuclei with pairing included, two-quasiparticle 
configuration which is located at some excitation energy with respect of quasiparticle 
vacuum at zero frequency  gets  lower in energy with increasing  rotational  frequency 
and becomes a vacuum configuration at some frequency $\omega_{pair}$ (see Fig.\ 
\ref{compar}b).  This change in vacuum configuration leads to a paired band crossing 
(Refs.\ \cite{BF.79,BF.79-2,VDS.83,Szy-book}). The experimental data related to 
the evolution of the energies of quasiparticle states with frequency and the properties
of paired band crossings (the frequency of crossing and the gain in alignment at the crossing) in 
rotational bands are, in general, well described in cranking model with quasiparticle 
energies defined in rotating frame (see, for example, Refs.\ \cite{BF.79,BF.79-2,GBDFH.94,ER.94,AER.00} 
for non-relativistic results and Refs.\  \cite{CRHB,AF.05,AO.13,A.14} for relativistic ones). 
Unpaired band crossings emerge from the crossing of two single-particle orbitals with 
the same quantum numbers which are separated by some energy at zero frequency/spin 
(see Fig.\ \ref{compar}c and Refs.\ \cite{NilRag-book,PhysRep-SBT}).  Again the  cranking 
models with single-particle orbitals, defined in rotating frame, well describe alignment 
properties  of the single-particle orbitals (which define the change of single-particle 
energies with frequency/spin) extracted from experimental data (see Refs.\  \cite{Rag.93,ALR.98,AF.05})
and experimental data on unpaired band crossings (see, for example, Refs.\  \cite{AKR.96,PhysRep-SBT}).

 These results clearly indicate that the cranking description  in rotating frame well reproduces 
the change of the energies of the single-particle states with rotational frequency.  Good agreement
with experiment  obtained in the cranking model proves that such a description is a good 
approximation to the description in the laboratory frame. This analysis also suggests that
that the transition from resonance to particle-bound part of rotational band takes
place at approximately the same frequency in rotating and laboratory frames. Note also
that the description of this transition is simpler than the description of band crossings
because the single-particle orbital does not interact with continuum threshold.
On the contrary, there is an interaction of the single-particle states at and near band crossing 
resulting in the repulsion of the single-particle orbitals (see Fig.\ \ref{compar}b,c). It is also important 
to mention the equivalence of the description of rotating nuclei in the intrinsic (rotating) and 
laboratory frames  was studied in  Ref.\ \cite{CEMPRRZ.95} on the example of $^{48}$Cr.

\begin{figure}[ht]
\centering
\includegraphics[width=4.0cm]{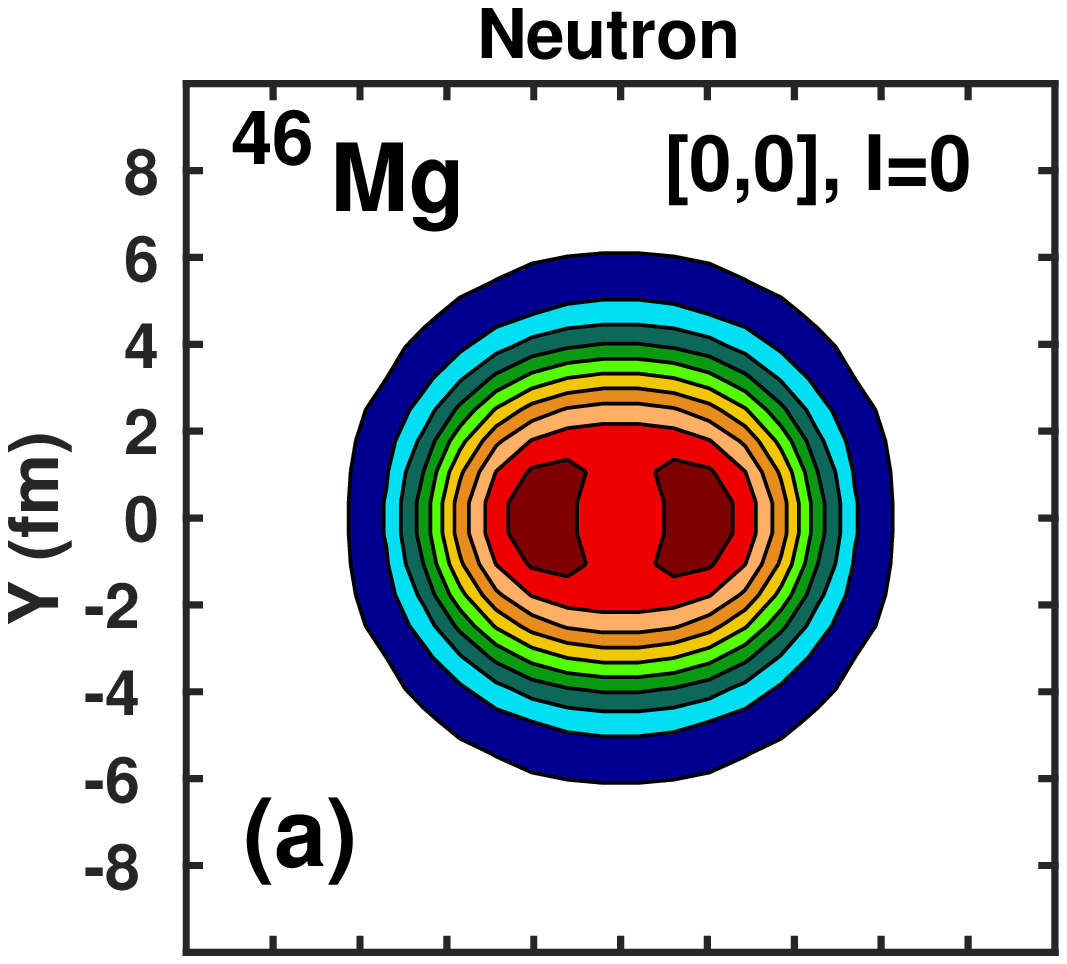}
\includegraphics[width=4.35cm]{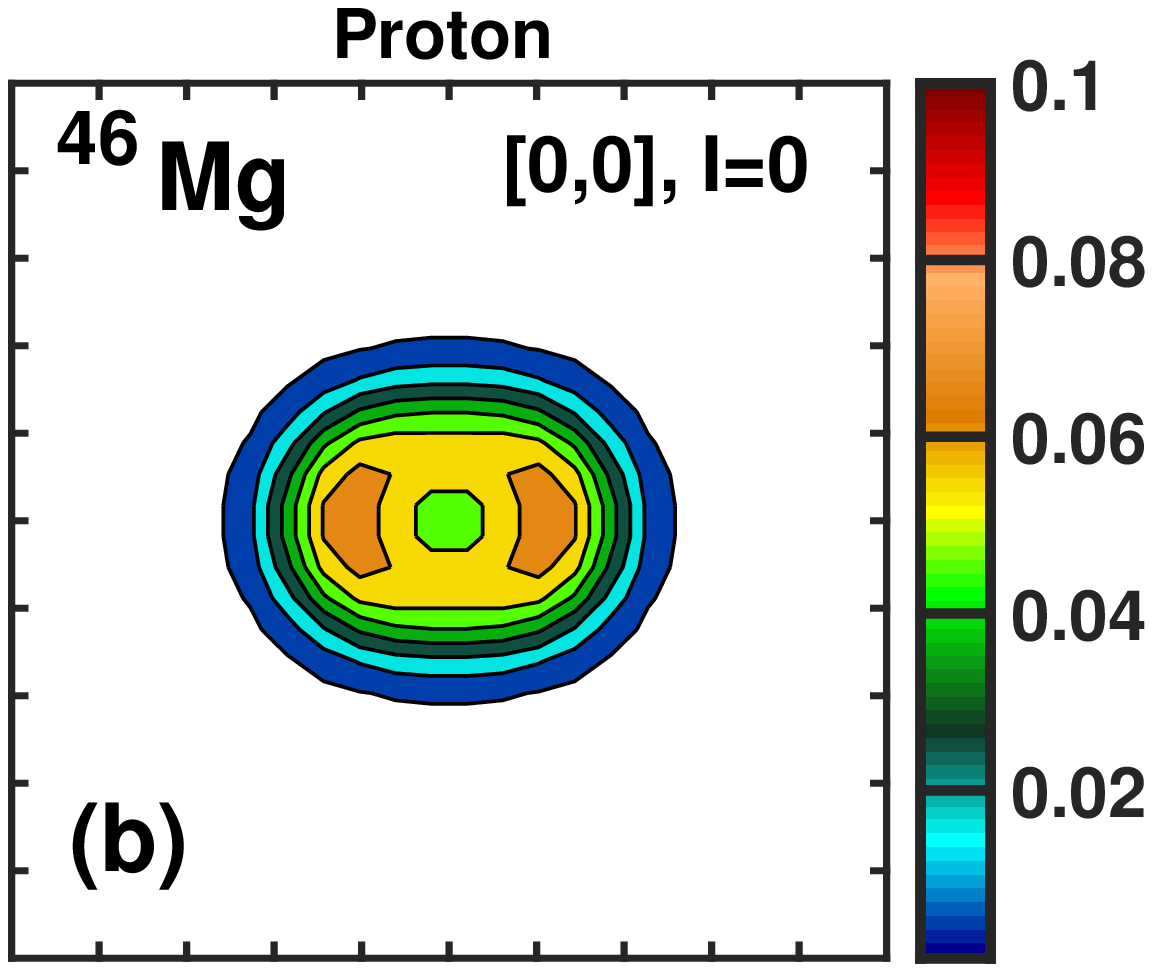}
\includegraphics[width=4.0cm]{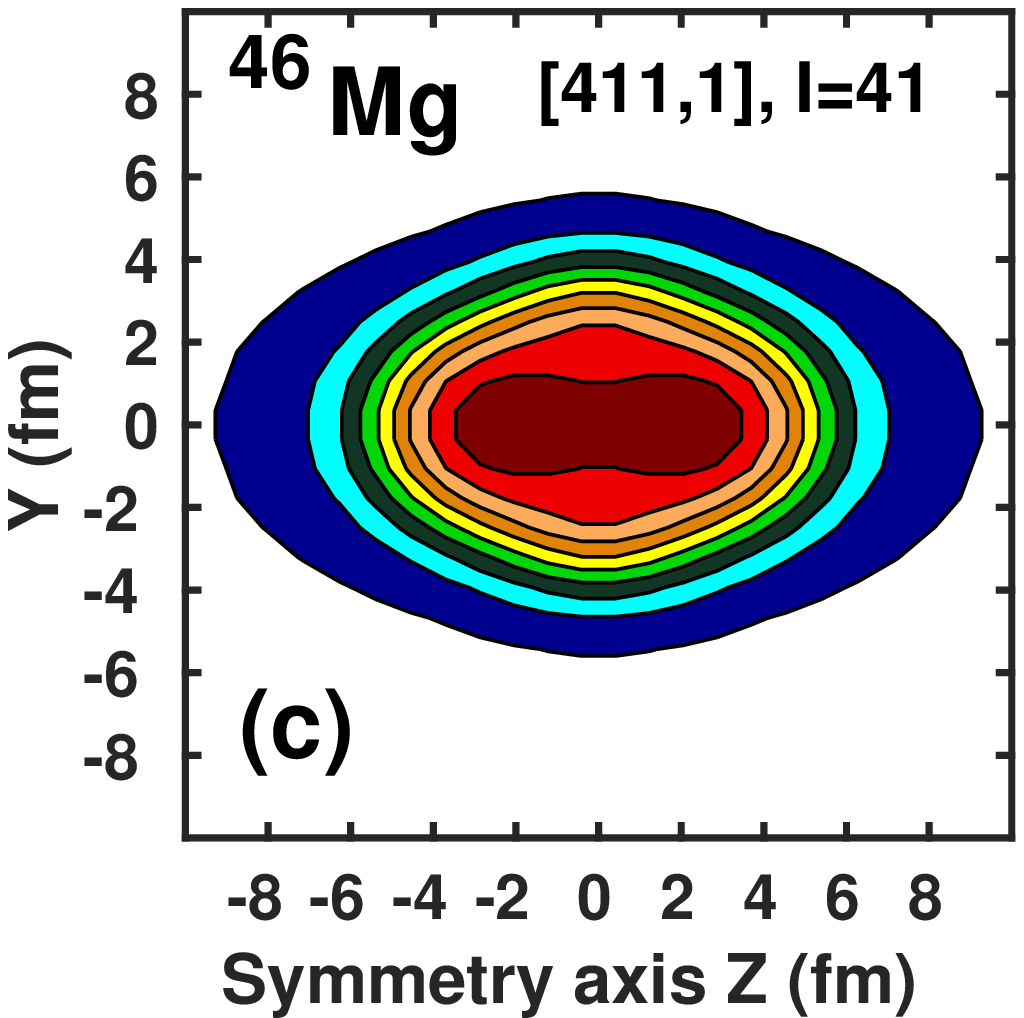}
\includegraphics[width=4.33cm]{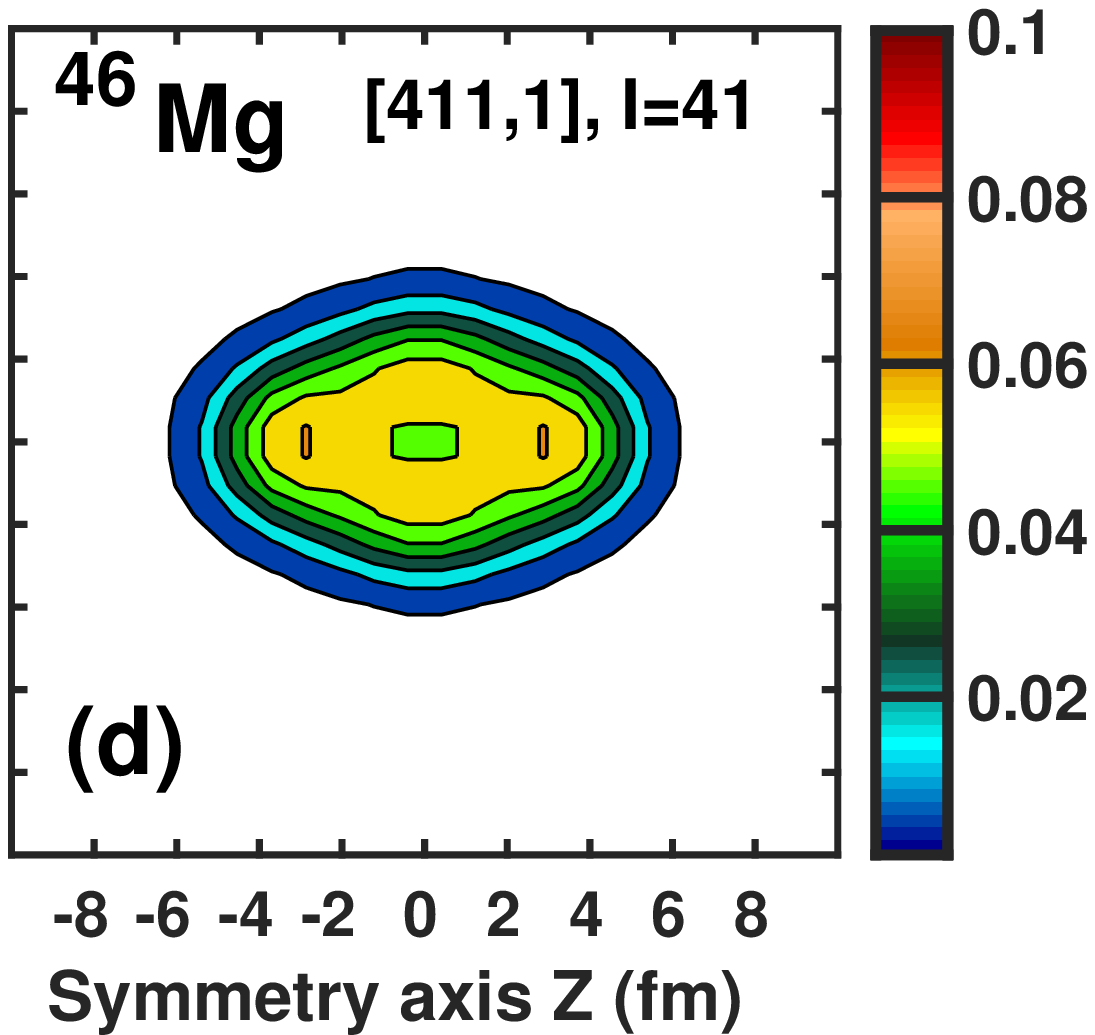}
\caption{(Color online) Neutron and proton density distributions of the 
[0,0] and [411,1] configurations in $^{46}$Mg at spins $I=0\hbar$ and 
$I=41\hbar$.  The density colormap starts at $\rho =0.005$ fm${^{-3}}$
and shows the  densities in fm$^{{-3}}$. Note that at these spins these 
configurations are nearly axially symmetric with $\gamma$-deformation
equal to $0^{\circ}$ and $\approx 3^{\circ}$ for the [0,0] and [411,1] configurations,
respectively.}
\label{density}
\end{figure}

  At highest spin displayed in Fig.\ \ref{E_RLD}, the configuration [411,1] is yrast. The routhian diagram 
of the  combined [5,1]/[411,1] configuration is shown in  Fig.\ \ref{routh}b. At low rotational frequency, 
three occupied neutron orbitals from the $N=4$ shell are located 
in continuum; this corresponds to the [5,1] configuration. With 
increasing rotational frequency these orbitals become particle-bound 
at  $\Omega_x \sim 1.6$ MeV. Thus, for the spins up to/beyond $I\sim 27.9\hbar$,
the [5,1] configuration represents the resonance/particle-bound band. At even 
higher  frequency of $\Omega_x \sim 1.8$ MeV ($I\sim 32.7\hbar$), the lowest 
$N=5$ and $N=6$ orbitals become occupied leading to the [411,1] configuration, 
which corresponds to particle-bound rotational band.
   
Fig.\ \ref{convergence}  illustrates the dependence of binding energies $E$ and
neutron single-particle energies $e_i$ on the number of fermionic shells $N_F$ 
used in the calculations. One can see that with increasing $N_F$, the configuration
[3,0] becomes more bound but each subsequent increase of $N_F$ leads to the 
decrease in the gain of the  binding energy so that the $N_F=18$ and $N_F=20$ results
differ by only $\approx 150$ keV; this difference represents 0.06\% of total binding
energy. Note that the slope and the curvature of $E-E_{RLD}$ curve defines
the moment of inertia (see Ref.\ \cite{PhysRep-SBT}); the similarities of these curves
calculated with different values of $N_F$ indicate that the moments of inertia
only weakly depend on $N_F$. Indeed, the difference of the moments of inertia
calculated with $N_F=18$ and $N_F=20$ is of the order of 0.6\%. The frequency
$\omega_{tran}$ at which the occupied orbital (the $3/2[431](r=-i)$ orbital in 
the [3,0] configuration) crosses the continuum threshold is another important quantity. 
This is because it defines the frequency (spin) at which the transition from resonance 
to particle-bound part of rotational band takes place.  Indeed, there is some dependence
of $\omega_{tran}$ on $N_F$ at low values of $N_F$;  $\omega_{tran} = 1.27$, 
1.17, 1.11 and 1.05 MeV at $N_F=10$, 12, 14 and 16, respectively. However, further 
increase of $N_F$ does not change $\omega_{tran}$. These features can be understood
from Fig.\ \ref{convergence}b. The single-particle energies of bound states are the
same at $N_F=18$ and $N_F=20$. However, the decrease of $N_F$ leads to some
minor changes in the energies of the single-particle states. Since the accuracy of
the description of the energies of the single-particle routhians is the same as for 
the states at no rotation, the crossing frequency $\omega_{tran}$ is well defined
by the bound parts of occupied routhians in the $N_F=18$ calculations.

  The situation  is somewhat different for the single-particle states in the continuum.
The energies of some of the states almost do not change with the increase of $N_F$ 
above  16. These are the 3/2[301], 3/2[312], 7/2[303], and 3/2[431] states
(see Fig.\ \ref{convergence}b). On the contrary,
although the energies of the 1/2[301] and 1/2[431] states show the trend to stabilization
with increasing $N_F$, they do not completely stabilize at $N_F=20$
(see Fig.\ \ref{convergence}b). The extrapolation 
to higher $N_F$ suggests the modification of their energies in the range of few hundred
keV as compared with the $N_F=20$ results. If these states are occupied in the configurations 
of interest, they may somewhat  modify (as compared with the solution in larger size basises) 
the energies of resonance parts of the bands. However, these modifications are rather modest 
(above mentioned few hundred keV) since the rotational properties of the configurations are 
predominantly defined by bound states. Note also that among considered configurations 
only few of them involve the occupation of the 1/2[431] orbital in the continuum.

  Fig.\ \ref{density} compares proton and neutron density distributions 
obtained for the ground state [0,0] (at spin $I=0$) and excited [411,1] (at 
spin $I=41\hbar$) configurations.  One can see substantial differences in the 
density distributions between these configurations and between proton and 
neutron density distributions within the same configuration. The later feature 
is due to the excess of neutrons over protons which also leads to the creation 
of neutron skin and larger and more spread out neutron density distribution as 
compared with proton  one.  The transition from the [0,0] to [411,1] configuration 
leads to a substantial elongation of density  distributions and to a larger neutron
skin in the axial direction. The density distributions of the configurations 
forming the yrast line in $^{46}$Mg are located in between of those shown 
in Fig.\  \ref{density} and follow the simple rule: the elongation of the
density distribution increases with the increase of intruder/hyperintruder
content of the configuration.  
   
    Fig.\ \ref{systematics} shows the results of the CRMF calculations for 
 even-even $^{42-50}$Mg nuclei. The first three nuclei are bound at spin zero 
 in the RHB calculations of Ref.\ \cite{AARR.14}, while the $^{48,50}$Mg 
 nuclei are unbound. The upper panels show both resonance and  particle-bound 
 parts of rotational bands. Note, that many calculated resonance bands may not exist 
 because of open decay channels. The bottom panels of Fig.\ \ref{systematics} show
 only particle-bound parts of rotational bands. The ground state bands
 of the [0,0] structure in  spin zero bound $^{42-46}$Mg nuclei are particle-bound in full 
 calculated spin range. On the contrary,  they consist mostly of resonance parts in 
 spin zero particle unbound $^{48,50}$Mg nuclei;  only two states at spin $I=4$ and 
 6 are particle-bound in the $^{48}$Mg nucleus and no such states exist in the [0,0]
 band of $^{50}$Mg. The rotational structures built on particle-hole excitations
 in the nuclei under study consist both of resonance and particle-bound parts.
 Although the balance of these two parts  depends on nucleus and configuration,
 the general trend of decreasing spin range of particle-bound parts of rotational
 bands with increasing neutron number are clearly seen.  Note that 
 resonance parts of rotational bands exists also in bound nuclei 
 located  below  the $I=0$ two-neutron drip line. 

   Fig.\ \ref{systematics} reveals for the first time new mechanism of
the extension  of nuclear landscape towards higher neutron numbers beyond
the $I=0$ two-neutron drip line. The $^{46}$Mg is last bound nucleus
at spin zero. However, particle-bound rotational bands exist in more neutron
rich $^{48,50}$Mg nuclei. These are the [0,0], [2,0], [3,0] [41,0] and [411,1] 
configurations in $^{48}$Mg (Fig. \ref{systematics}i) and the [511,1] 
configuration in $^{50}$Mg (Fig. \ref{systematics}j). Particle-hole excitations 
building these configurations increase their intruder/hyperintruder content and 
deformations. The combined effect of these two factors changes their character 
from resonance type at low spin to particle-bound at high spin.  While this is
quite frequent effect in $^{48}$Mg, this transition, which takes place at very
high spin, requires significant deformation and high intruder/hyperintruder 
content in the $^{50}$Mg nucleus which is located four neutrons beyond the 
$I=0$ two-neutron drip line.
    
   Note that this mechanism has large similarities with the one seen recently
in   proton-rich hyperheavy $(Z\geq126$) nuclei (see discussion in Sect. IX 
of Ref.\  \cite{AATG.19}).  Proton chemical potential $\lambda_\pi$ has 
a pronounced slope as a function of quadrupole deformation $\beta_2$: for 
a given nucleus the proton Fermi level is located at substantially lower 
energies at toroidal shapes as compared with ellipsoidal ones. This leads 
to the transition from proton-unbound ellipsoidal nuclear shapes to proton-bound 
toroidal ones. In a similar way in rotating nuclei the transition
from unbound (resonance) to bound part of rotational band is triggered
by an increase of other collective coordinate, namely, rotational 
frequency. Similar to rotating nuclei, above mentioned feature of $\lambda_\pi$ in 
hyperheavy nuclei is a  source of unusual shift in the position of two 
proton-drip line toward more proton rich nuclei (as compared with 
general trend seen in the $(Z,N)$ plane for the $Z<120$ nuclei) (see
Ref.\ \cite{AATG.19}).
 
  The current investigation neglects the pairing correlations.  While this is 
crude approximation for the ground state bands, it becomes much more realistic with
increasing of spin and the number of particle-holes excitations involved in
the building of nucleonic configurations. Even if some minor pairing correlations will
remain active in the high-spin configurations, the general features will not be strongly
affected. This is because the neutron Fermi level will be located in the vicinity 
of last occupied neutron orbital (as defined in the calculations without pairing),
and, thus it will also be diving into the region of negative energies with 
increasing rotational frequency.

   The possibility to observe particle-bound parts of rotational bands by means
of traditional gamma-spectroscopy depends on the competition of gamma-decay 
with neutron(s)  emission. For such an observation,  the half-lives of gamma-decays 
from the states within the rotational bands have to be shorter than the half-lives of 
neutron(s) emission. Gamma-decay half-lives are expected to be similar to the 
ones of experimentally known rotational bands; typically these half-lives are 
extremely short being in nanosecond ($10^{-9}$ s) range.  On the contrary, 
neutron(s) emission from rotational states of particle-bound parts of rotational 
bands has not been studied so far. 
In general, the suppression of the coupling of the bound states with continuum
due to decreasing role of  pairing with increasing spin is expected to lead to a
substantial  suppression of neutron emission at high spin. Note that weakly 
bound states along the proton drip line still form rotational bands which are 
seen in experiment. It is reasonable to expect that this will also be true for 
bound parts of rotational bands of near-neutron drip line nuclei. However, a 
quantitative answer on the question of the competition of gamma- and neutron(s) 
decays requires an additional investigation which will be performed in future. Note 
that the existence of particle-bound rotational bands will not be affected even in the 
case when neutron(s)-decay rates are shorter than gamma-decays; this will simply 
change the tool of their observation from gamma-spectroscopy to 
particle-spectroscopy.  With nuclear timescale of $\tau \sim 10^{-23}$ s, there is 
huge range of half-lives in which the particle-bound rotational states can exist. For 
example, in nuclear world the analog of the Earth (extremely stable object inhabited 
by humans) would have existed only for time of $\sim 4.5 \cdot 10^{-14}$ sec = $\tau$ 
times $4.5 \cdot 10^9$ revolutions of Earth around Sun. This time is by orders of 
magnitude shorter than the  typical gamma-decay half-lives of rotational states.

  In conclusion, the investigation of rotational properties of the Mg isotopes 
in the vicinity of two-neutron drip line has been performed in the cranked
relativistic mean field theory. It reveals two new physical mechanisms
active in the vicinity of neutron drip line which have not been discussed
before in the literature. Nucleonic configurations having occupied
neutron single-particle orbitals in continuum lead to the formation of 
resonance bands in the calculations; many of them will not exist 
in nature because of open decay channels. However, occupied 
intruder/hyperinitruder orbitals of these configurations initially located in 
continuum dive (because of their high-$j$ content leading to strong
Coriolis force) into the region of negative 
energies  with increasing rotational frequency and become particle-bound. This  
mechanism  leads  to the birth of particle-bound rotational bands. Alternative 
possibility of the transition from particle-bound to resonance part of rotational 
bands (the death of particle-bound rotational bands) with increasing spin also 
exists but it is less frequent since it requires the occupation of the orbital, 
strongly up-sloping in rotational frequency, which raises from negative to 
positive energy with increasing rotational frequency. These features of the 
birth of particle-bound rotational bands provide a new mechanism of the 
extension of nuclear landscape  at non-zero spin to neutron numbers which 
are larger than those for two-neutron drip line at spin zero.

\section*{Acknowledgements}

This material is based upon work supported by the U.S. Department of Energy,  
Office of Science, Office of Nuclear Physics under Award No. DE-SC0013037
and KAKENHI Grant No. 17K05440.

\section*{References}

\bibliographystyle{physlett.bst}
\bibliography{article-plb-final.bib}

\end{document}